\documentclass[11pt,a4paper]{article}
\usepackage{amsmath,amssymb}
\usepackage{epsfig,graphicx}
\usepackage[english]{babel}
\usepackage{verbatim}
\usepackage{cite}

\def\starup#1{\mbox{$\raise1.6ex\hbox{$*$} \kern-0.5em#1$}}
\def\krup#1{\mbox{$\raise1.8ex\hbox{$+$} \kern-1.0em#1$}}

\begin{document}

\title{\bf Cross Section and Forward-Backward Asymmetry of $t\bar{t}$ Production in the Model with Four Color Symmetry}
\author{M.~V.~Martynov$^a$\footnote{{\bf e-mail}: martmix@mail.ru},
A.~D.~Smirnov$^{a}$\footnote{{\bf e-mail}: asmirnov@uniyar.ac.ru}
\\
$^a$ {\small Division of Theoretical Physics, Department of Physics,}\\
{\small Yaroslavl State University, Sovietskaya 14,}\\
{\small 150000 Yaroslavl, Russia.}}
\date{}
\maketitle

\begin{abstract}
 The contributions
to the cross section $\sigma_{t\bar{t}}$ and to the forward-backward asymmetry
$A_{\rm FB}^{t \bar t}$ of $t\bar{t}$ production at the Tevatron from  $Z'$-boson and scalar leptoquarks $S^{(\pm)}_a$ and scalar gluons $F_a$  predicted by the minimal model with four color quark-lepton symmetry
are calculated.
These contributions  are shown to be small in tree approximation  and can be significant with account of the 1-loop $gt\bar{t}$ effective vertex induced by the scalar doublets.
The lower mass limit for scalar gluons  $m_F \gtrsim 320\, GeV$  from the Tevatron data  is obtained and it is shown
that for $m_{F_1}\lesssim 990\,GeV$ the scalar gluon $F_1$ can be evident at LHC at the significance not less that $3\sigma$ (for $\sqrt{s}=14$ TeV, $L=10\,fb^{-1}$).
\end{abstract}

\section{Introduction. Minimal Quark-Lepton Symmetry model.}


The search for a new physics beyond the Standard Model (SM) is now one
of the aims of the high energy physics.
One of the
new physics can be
induced by the
possible  four color symmetry treating leptons as quarks
of the fourth color \cite{pati_salam}.
This symmetry can be unified with the SM by the gauge group
\begin{eqnarray}
     G_{new}=G_c \times SU_L(2) \times U_R(1)
\label{eq:Gnew}
\end{eqnarray}
where $G_c$ is the group of the four color symmetry.
The color group $G_c$ can be the vectorlike group
$G_c = SU_V(4)$ or the general chiral group $G_c = SU_L(4) \times SU_R(4)$
or  one of the special groups of the left or right four color symmetry
$G_c = SU_L(4) \times SU_R(3)$, $G_c = SU_L(3) \times SU_R(4)$.

%
%
%
%


The Minimal four color Quark-Lepton Symmetry model (MQLS-model)
is based on the gauge group
\begin{eqnarray}
G_{new}=SU_{V}(4)\times SU_{L}(2)\times U_{R}(1)
\label{eq:G421}
\end{eqnarray}
as on the minimal group containing the four color
symmetry of quarks and leptons~\cite{AD22,smirnov-1995-346}.

According to this group   in addition to
gluons $G_\mu^j$, $j = 1, 2, \ldots, 8$ and
$W^\pm$-, $Z$-bosons the  gauge sector
predicts the  new gauge particles: vector leptoquarks
$V_{\alpha \mu}^\pm$, $\alpha = 1, 2, 3$
with charges $Q_V^{em} = \pm 2/3$
and an  extra Z'-boson originating from the four color quark-lepton symmetry.

\underline{Fermion sector of the model}

In MQLS-model quarks and leptons form the $SU_V(4)$-quartets $\psi_{p a A}$,
$A$ = 1, 2, 3, 4,  $a$ = 1, 2, $p$ = 1, 2, 3, $\ldots$
\begin{displaymath}
\psi'_{p 1 A} \; : \quad
\left ( \begin{array}{c} u'_\alpha \\ \nu'_e \end{array} \right ) , \;
\left ( \begin{array}{c} c'_\alpha \\ \nu'_\mu \end{array} \right ) , \;
\left ( \begin{array}{c} t'_\alpha \\ \nu'_\tau \end{array} \right ) , \;
\cdots
\end{displaymath}
\begin{displaymath}
\psi'_{p 2 A} \; : \quad
\left ( \begin{array}{c} d'_\alpha \\ {e^-}' \end{array} \right ) , \;
\left ( \begin{array}{c} s'_\alpha \\ {\mu^-}' \end{array} \right ) , \;
\left ( \begin{array}{c} b'_\alpha \\ {\tau^-}' \end{array} \right ) , \;
\cdots
\end{displaymath}
\noindent
where
${Q'}^{L,R}_{p a \alpha}$,~${\ell'}^{L,R}_{p a}$ are the basic left and right quark and lepton fields.

Each lepton have $SU_V(4)$ "color" $A=4$.

\underline{Fermion mixing in MQLS.}

The basic left and right quark and lepton fields
${Q'}^{L,R}_{p a \alpha}$,
${\ell'}^{L,R}_{p a}$
can be written, in general, as superpositions
\begin{eqnarray}
{Q'}^{L,R}_{p a \alpha} = \sum_q \left ( A^{L,R}_{Q_a} \right )_{p q}
Q^{L,R}_{q a \alpha} ,   \qquad
{l'}^{L,R}_{p a} = \sum_q \left ( A^{L,R}_{l_a} \right )_{p q}
l^{L,R}_{q a},
\label{eq:mixf}
\end{eqnarray}

\noindent of mass eigenstates
$Q^{L,R}_{q a \alpha}$, $\ell^{L,R}_{q a}$. Here
$A^{L,R}_{Q_a}$ and $A^{L,R}_{\ell_a}$ are unitary matrices
diagonalizing the mass matrices of quarks and leptons respectively.

$(A^L_{Q_1})^+ A^L_{Q_2}\equiv C_Q = V_{CKM}$ is Cabibbo-Kobayashi-Maskawa matrix
$(A^L_{\ell_1})^+~A^L_{\ell_2}~\equiv~ C_\ell$ is the analogous lepton mixing matrix~{\small($(C_l)^+~=~U_{PMNS}$)}

$(A^{L,R}_{Q_a})^+ A^{L,R}_{\ell_a}\equiv K^{L,R}_a$ are the new mixing matrices
which are specific for the models with the four color symmetry.


\underline{The interaction of the gauge fields with the fermions} has the form
\begin{eqnarray}
{\cal L}_\psi^{gauge}
=
{\cal L}_\psi^{V} + {\cal L}_\psi^{W} +
{\cal L}_\psi^{QCD} +  {\cal L}_\psi^{QED} +
{\cal L}_\psi^{NC},
\label{eq:Lint}
\end{eqnarray}
where
\begin{eqnarray}
{\cal L}_\psi^{V}
& = & \frac{g_4}{\sqrt 2} \big\lbrace
\big ( \bar Q_{p a \alpha} \big [  \big ( K^L_a \big )_{p q} \gamma^\mu P_L +
  \big ( K^R_a  \big )_{p q} \gamma^\mu P_R \big ]
\ell_{q a} \big ) V^\alpha_\mu + h.c. \big\rbrace
\label{eq:LintV} , \\
{\cal L}_\psi^{W}
& = & \frac{g_2}{\sqrt 2}   \big\lbrace   \big [
\bar Q_{p 1 \alpha}  \big ( C_Q \big )_{p q} \gamma^\mu P_L Q_{q 2 \alpha}  +
\bar \ell_{p 1}  \big ( C_\ell \big )_{p q}\gamma^\mu P_L \ell_{q 2}
\big ] W^+_\mu
+   h.c. \big\rbrace,
 \label{eq:LintW} \\
{\cal L}_\psi^{QCD} & = & g_{st} G_\mu^j \big ( \bar Q \gamma^\mu t_j Q \big ),
 \label{eq:LintQCD} \\
{\cal L}_\psi^{QED} & = & - |e| A_\mu \big ( \bar \psi \gamma^\mu Q^{em} \psi \big ),
 \label{eq:LintQED} \\
%
{\cal L}_\psi^{NC} & = & - Z_\mu J^Z_{\mu} - Z'_\mu J^{Z'}_{\mu} .
\label{eq:Lnc}
%
\end{eqnarray}


%
%
%
%

\underline{Features of $Z'$-boson originating from the four color symmetry.}

In general case the mass eigenstates $Z$ and $Z'$ are superposition
of two basic fields $Z_1$ and $Z_2$.
In MQLS model the $Z-Z'$ mixing angle is small ($\theta_m < 0.006$) and
we neglect below the $Z-Z'$ mixing believing $Z \approx Z_1$ and $Z' \approx Z_2$.
The interaction of the neutral gauge fields with the fermions has the form
\begin{equation}
{\cal L}_{\rm N C}^{\rm gauge} =  -e Z_{1\mu} J^{Z_1}_{\mu} -
\frac{e}{c_W}Z_{2\mu} J^{Z_2}_{\mu} , \label{one}
\end{equation}
 \noindent
\begin{eqnarray}
J^{Z_1}_\mu &=& \bar f \gamma_\mu (v_f^{Z_1}+a_f^{Z_1}\gamma_5)f,
\\ \nonumber J^{Z_2}_\mu &=& \bar f \gamma_\mu
(v_f^{Z_2}+a_f^{Z_2}\gamma_5)f, \label{two}
\end{eqnarray}
with couplings
\begin{eqnarray}
v_{f_a}^{Z_2}&=&\frac{1}{s_S\sqrt{1-s_W^2-s_S^2}}\left[
c_W^2\sqrt{\frac{2}{3}}(t_{15})_f-
\left(Q_{f_a}-\frac{(\tau_3)_{aa}}{4}\right)s_S^2\right], \,\label{three}\\
a_{f_a}^{Z_2}&=&
\frac{s_S}{\sqrt{1-s_W^2-s_S^2}}\frac{(\tau_3)_{aa}}{4}.
\label{fourth}
\end{eqnarray}

The fermionic decays of $Z'$ boson
are defined by the coupling constants~(\ref{fourth}) and
the corresponding partial widths of $Z'$ boson decays to $f_a \bar{f_a}$ pairs
for $m_{f_a} \ll M_{Z'}$ have the form \cite{Smirnov:2009ed}
\begin{equation}
\label{GamToFermion}
\Gamma(Z'\rightarrow f_a \bar{f_a})=N_f M_{Z'}\frac{\alpha}{3} \big((v_{f_a}^{Z'})^2+(a_{f_a}^{Z'})^2 \big),
\end{equation}
where the color factor $N_f=1(3)$ for leptons(quarks) $f=l(q)$.

Writing the interaction of $Z'$ boson with scalar field $\Phi$ as
\begin{equation}
\mathcal{L}_{Z' \Phi \Phi}= i g_{Z' \Phi \Phi} Z'_{\mu}
\left(\partial^{\mu}\Phi^{*}\Phi - \Phi^{*} \partial^{\mu}\Phi
\right),
\end{equation}
where $g_{Z' \Phi \Phi}$ is the corresponding coupling constant
for the width of $Z'$ boson decay into $\Phi \tilde{\Phi}$ pair we have the expression
\begin{equation}
\label{ScalarPartialWidth}
\Gamma (Z' \rightarrow \Phi \tilde{\Phi})= N_{\Phi} M_{Z'} \frac{  g_{Z'\Phi \Phi}^2 }{48\pi}
\left(1-\frac{4 m_{\Phi}^2}{M^2_{Z'}}
\right)^{3/2}
\end{equation}
where $N_{\Phi}$ is the color factor ($N_{F_a}=8$ for scalar gluons, $N_{S^{(\pm)}_a}=3$ for
scalar leptoquarks, $N_{\Phi_a'}=1$ for the additional colorless scalar doublet)
and $m_{\Phi}$ is a mass of the scalar particle.

The scalar gluons $F_{a}$ and the scalar leptoquarks $S^{(\pm)}_{a}$ gives the main
contribution into $Z'$ boson width of type~(\ref{ScalarPartialWidth}).
The coupling constants of these particles with $Z'$ boson are predicted by the MQLS model as
\begin{align}
\label{couplings}
g_{Z'F_aF_a}=-\frac{e}{2}\frac{\sigma}{s_W c_W},\quad
g_{Z'S^{(\pm)}_a S^{(\pm)}_a}=-e\left(\frac{\sigma}{2s_Wc_W}\pm\frac{2 t_W}{3\sigma}\right),
\end{align}
where  $t_W = \tan \theta_W$ and  $\sigma=s_W s_S/\sqrt{1-s_W^2-s_S^2}$.

The parameter $s_S$ is defined by the mass scale $M_c \sim M_V$ of the four-color symmetry breaking
and by the intermediate mass scale $M'\sim M_{Z'}$.
For example for $M' \sim  10 \, TeV$ and for $M_c = 10^4 \, TeV, \, 10^6 \, TeV, \, 10^8 \, TeV$
we have $s_s^2 = 0.070, \, 0.112, \, 0.154$ respectively~\cite{AD22}.
For numerical estimations we use below the value $s_s^2=0.114$ which corresponds to
$M_{Z'} \sim 1 - 5 \, TeV$ and $M_c \sim 10^3 \, TeV$.
With these values of $s_s^2$ and of the masses of scalar particles
the relative total width of $Z'-$boson $\Gamma_{Z'}/M_{Z'}$ occurs to be equal  to
\begin{equation}
\label{gammaZ1}
\Gamma_{Z'}/M_{Z'} = 4.3 \, \% \; (1.1 \, \% , \;  3.2 \, \% ), \;\; 5.2 \,\% \; (2.0 \,\% , \;  3.2 \,\% ),
\;\;  5.3 \,\% \; (2.1 \,\% , \;  3.2 \,\%)
\end{equation}
for $M_{Z'}$ of about respectively 1 TeV, \, 3 TeV, \, 5 TeV and above,
the corresponding values of the relative widths
of the $Z'$~decays respectively into scalar particles and into fermions are shown in parenthesis.

%
%


\underline{The scalar sector} contains in general four multiplets
\cite{smirnov-1995-346,AD22,povarov-smirnov-2001}
\begin{eqnarray}
(4,1,1):   \Phi^{(1)} & =  &
\left ( \begin{array}{c} S_{\alpha}^{(1)}    \\
  \frac {\eta_1 + \chi^{(1)}+i\omega^{(1)}}{\sqrt 2} \end{array} \right ) ,
\nonumber \\
(1,2,1):   \Phi_a^{(2)} & = & \delta_{a2}\frac{\eta_2}{\sqrt 2} + \phi_a^{(2)},
\nonumber \\
\mathbf{(15,2,1):}  \Phi_a^{(3)} & = & \left(
\begin{array}{cc}
 \mathbf{(F_a)_{\alpha\beta}} & \mathbf{S_{a\alpha}^{(+)}} \\
  \mathbf{S_{a\alpha}^{(-)}}  &  0
\end{array} \right) +  \Phi_{15,a}^{(3)}t_{15},
 \\
(15,1,0):  \Phi^{(4)} & = & \left(
\begin{array}{cc}
 F_{\alpha\beta}^{(4)} & \frac 1{\sqrt 2}S_{\alpha}^{(4)} \\
\starup{S_\alpha^{(4)}}   &  0
\end{array} \right) + (\eta_4 + \chi^{(4)})t_{15} ,
\nonumber
\end{eqnarray}
transforming according to the  (4,1,1)-,(1,2,1)-,(15,2,1)-,(15,1,0)-
representations of the
$SU_V(4) \times SU_L(2)\times U_R(1)$-group
respectively. Here
$
 \Phi_{15,a}^{(3)}= \delta_{a2}\eta_3 + \phi_{15,a}^{(3)},
$
$\eta_1$, $\eta_2$, $\eta_3$, $\eta_4$ are the vacuum expectation
values.

Third multiplet $(15.2.1)$ interacts with quarks.
\begin{eqnarray}
(15.2.1):\Phi^{(3)}: \quad \,\,\,\,\,\,\,
\left ( \begin{array}{c} S_{1\alpha}^{(+)}  \\
         S_{2\alpha}^{(+)}\end{array} \right );
\left ( \begin{array}{c} S_{1\alpha}^{(-)}  \\
         S_{2\alpha}^{(-)}\end{array} \right );
\left ( \begin{array}{c} F_{1k}  \\
         F_{2k}\end{array} \right );
\left ( \begin{array}{c} \Phi_{1,15}^{(3)}  \\
         \Phi_{1,15}^{(3)}\end{array} \right ),
\label{eq:a2}
\end{eqnarray}
where $\mathbf{S^{(\pm)}_{a\alpha}}$ and $\mathbf{F_{ak}}$ (k=1,2...8) are the scalar
leptoquark and scalar gluons doublets.
$\Phi^{(3)}_{15}-\Phi^{(2)}$-mixing gives the SM Higgs doublet $\Phi^{(SM)}$
 and an additional $\Phi'$ doublet.
These scalar doublets have the electric
charges
\begin{eqnarray}
Q_{em}:\quad \,\,\,\,\,\,\,
\left ( \begin{array}{c} 5/3  \\
         2/3\end{array} \right );
\left ( \begin{array}{c} 1/3  \\
        - 2/3\end{array} \right );
\left ( \begin{array}{c} 1  \\
         0 \end{array} \right );
\left ( \begin{array}{c} 1  \\
         0 \end{array} \right ).
\nonumber
\end{eqnarray}

In general
\begin{eqnarray}
S_{2 \alpha}^{(+)}&=&\sum_{m=0}^3 c_m^{(+)}S_m, \; \; \; \; \; \;
 \starup{S_2^{(-)}}=\sum_{m=0}^3 c_m^{(-)}S_m,
\nonumber
\label{eq:mixS}
\end{eqnarray}
where $S_m$ are three physical leptoquarks  with
electric charge 2/3 and $S_0$ is the Goldstone mode,
 $c^{(\pm)}_{m}$ are the elements of the
unitary scalar leptoquark mixing matrix, $|c^{(\pm)}_{0}|^2=\frac
{1} {3}g_4^2 \eta_{3}^{2}/m_V^2 \ll 1$.

The experimental lower mass limits for the scalar leptoquarks from their direct search are
\cite{PDG2008}

\begin{eqnarray}
m_{LQ}&\gtrsim & 250\,  \, \mbox{GeV}.
\end{eqnarray}

The indirect data set the limits on the relations of  scalar leptoquark coupling constants to their masses.


In MQLS-model the leptoquark Yukawa coupling constants  are (due their Higgs origin) proportional to
the ratios $m_{f}/ \eta $ of the fermion masses $m_f$ to the SM
VEV $\eta$. As a result these coupling constants are known (up to mixing parameters) and are small for light quarks.
So, the indirect mass limits for MQLS scalar leptoquarks are weaker then those from direct searchers.

\underline{Mass limits for scalar gluons $F_{ak}$.}

\begin{figure}[htb]
\begin{center}
\includegraphics[scale=0.3,keepaspectratio=true]{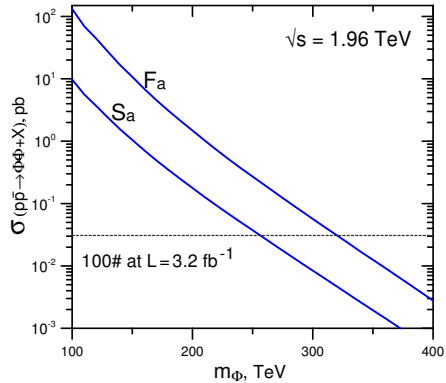}
\caption{Cross sections of $SS^*$-,$FF^*$-pair production at the Tevatron as functions of the masses of scalar particles.}
\end{center}
\end{figure}
The partonic cross sections of scalar gluon pair production are  known \cite{manohar-2006-74,Martynov:2008wf,martynov-smirnov-2010}, which gives now possibility to calculate cross
section of scalar gluon pair production at the Tevatron in dependence on scalar gluon mass.
In these calculations
we use  PDF's set AL'03 \cite{alekhin}~
(NLO,~variable-favor-number) with the  K-factor chosen as
 $K=1.45$ for consistency with theoretically predicted dependence
of $\sigma^{NLO}(t\bar{t})$ on $m_t$
\cite{Cacciari:2008zb,Kidonakis:2008mu}.

Our estimate for mass limits for scalar gluons $F_a$ from direct searches at Tevatron is
\begin{eqnarray}
m_{F_a}&\gtrsim &320\,  \, \mbox{GeV.}
\end{eqnarray}

\underline{Possibility of the direct searches scalar gluon at the LHC}

The production
cross section of scalar gluons $F$ at the LHC with masses $m_F \lesssim 1300\,
GeV$ is shown to be sufficient
for the effective ($N_{events}\gtrsim 100$) production  of these particles at the LHC ($L=10\,fb^{-1}$) \cite{Martynov:2008wf}.

At $m_{F_1}\lesssim 990\,GeV$ from analysis statistical significance the number of the signal $t\bar t b \bar b$ events will exceed the SM background by $3\sigma$ (LHC $L=10\,fb^{-1}$) \cite{martynov-smirnov-2010}.

\underline{The interaction of the fermions with the scalars.}

The Yukawa interaction of the fermions with the scalar $SU_L(2)$- doublets
$\phi^{(2)}$ and $\phi^{(3)}_i$ has, in general, the form
\begin{equation}
{\cal L}_\psi^{Yukawa} = - \bar \psi'^L_{p a A} \big [
\big ( h_b \big )_{p q} \phi^{(2) b}_a \delta_{A B} +
\big ( h'_b \big )_{p q} \phi^{(3) b}_{i a} \big ( t_i \big )_{A B}
\big ] \psi'^R_{q b B} + h.c. ,
\label{eq:LYukawa}
\end{equation}

\noindent where
$\phi^{(2) 2}_a = \phi^{(2)}_a$,
$\phi^{(2) 1}_a = \varepsilon_{a c} (\phi^{(2)}_c)^*$,
$\phi^{(3) 2}_{i a} = \phi^{(3)}_{i a}$,
$\phi^{(3) 1}_{i a} = \varepsilon_{a c} (\phi^{(3)}_{i c})^*$,
$i$ = 1, 2, $\ldots$, 15, $\varepsilon_{a c}$ is antisymmetrical symbol,
$h_b$ and $h'_b$ are four arbitrary matrices.

After symmetry breaking this Lagrangian
gives the arbitrary masses to the quarks and
leptons and gives the interactions of fermions with the scalar fields
\begin{eqnarray}
{\cal L}^{int}_\Psi&  = & {\cal L}_{\chi^{(SM)} ff} +
{\cal L}_{\Phi' ff} + {\cal L}_{FQQ} + {\cal L}_{SQl}.
\label{eq:lint}
\end{eqnarray}
$$ h \sim m_{f}/ \eta, $$
$$ m_u/ \eta \sim m_d/ \eta \sim 10^{-5},  m_s/ \eta \sim 10^{-3}, \,\,\,\,
m_c/ \eta \sim m_b/ \eta \sim 10^{-2}, \,\,\,\,$$
   $$  m_t/ \eta \sim 0.7. \,\,\,\, !!!   $$

The interactions of the scalar leptoquarks $S^{(\pm)}_{a\alpha}$ with quarks and leptons:
\begin{eqnarray}
L_{S^{(+)}_1 u_i l_j} &=& \bar u_{i\alpha }  \Big [ (
h^L_+)_{ij}P_L + (h^R_+)_{ij}P_R  \Big ] l_{j} S_{1\alpha}^{(+)} +
{\rm h.c.},
\nonumber\\
L_{S^{(-)}_1 \nu_i d_j} &=& \bar \nu_{i} \Big [ ( h^L_-)_{ij}P_L +
(h^R_-)_{ij}P_R  \Big ] d_{j\alpha } S_{1\alpha}^{(-)} + {\rm
h.c.},
\label{eq:Lsql} \\
L_{ S_m u_i \nu_j} &=& \bar u_{i\alpha} \Big [
(h^L_{1m})_{ij}P_L+(h^R_{1m})_{ij}P_R \Big ] \nu_{j} S_{m\alpha}
+{\rm h.c.}\nonumber
\\
L_{ S_m d_i l_j} &=& \bar d_{i\alpha} \Big [ (h^L_{2m})_{ij}P_L +
(h^R_{2m})_{ij}P_R \Big ] l_{j} S_{m\alpha} +{\rm h.c.} \nonumber
\end{eqnarray}

The interactions of the scalar gluons with quarks:
\begin{eqnarray}
L_{F_1 u_i d_j} &=& \bar u_{i\alpha }  \Big [ ( h^L_{F_1})_{ij}P_L
+ (h^R_{F_1})_{ij}P_R  \Big ] (t_k)_{\alpha\beta}d_{j\beta} F_{1k}
+ {\rm h.c.},\nonumber
\\
L_{ F_2 u_i u_j} &=& \bar u_{i\alpha} \Big [ ( h^L_{1F_2})_{ij}P_L
\Big ](t_k)_{\alpha\beta} u_{j\beta } F_{2k} + {\rm h.c.},
\\
L_{ F_2 d_i d_j} &=& \bar d_{i\alpha} \Big [ ( h^R_{2F_2})_{ij}P_R
\Big ](t_k)_{\alpha\beta} d_{j\beta } F_{2k} + {\rm h.c.}\nonumber
\end{eqnarray}

\underline{Scalar leptoquarks $S^{(\pm)}_1$, $S_m$ couplings to fermions:}

\begin{eqnarray}
(h^{L}_+)_{ij} &=&\sqrt{ 3/2} \frac{1}{\eta \sin\beta}
\Big [  m_{u_i} (K_1^LC_l)_{ij} -  (K^R_1)_{ik}m_{\nu_i} (C_l)_{kj} \Big ],
\nonumber \\
(h^{R}_+)_{ij} &=&-\sqrt{ 3/2} \frac{1}{\eta \sin\beta}
\Big [ (C_Q)_{ik}m_{d_k} (K_2^R)_{kj} - m_{l_j} (C_QK_2^L)_{ij}  \Big ],
\nonumber \\
(h^{L}_-)_{ij} &=&\sqrt{ 3/2} \frac{1}{\eta \sin\beta}
\Big [ (\krup{K^R_1})_{ik} m_{u_k} (C_Q)_{kj} - m_{\nu_j}
(\krup{K_1^L}C_Q)_{ij}  \Big ],\label{const1}
\\
(h^{R}_-)_{ij} &=&-\sqrt{ 3/2} \frac{1}{\eta \sin\beta}
\Big [( C_l \,\krup{K^L_2} )_{ij} m_{d_j}  -(C_l)_{ik} m_{l_k}
(\krup{K_2^R})_{kj}  \Big ], \nonumber\\
(h^{L,R}_{1m} )_{ij} &=& -\sqrt{ 3/2} \frac{1}{\eta \sin\beta}
\Big [  m_{u_i} (K_1^{L,R})_{ij} -
(K^{R,L}_1)_{ij}m_{\nu_j} \Big ] c_m^{(\pm)},
\nonumber\\
(h^{L,R}_{2m})_{ij} &=&-\sqrt{ 3/2} \frac{1}{\eta \sin\beta}
\Big [  m_{d_i} (K_2^{L,R})_{ij} -
(K^{R,L}_2)_{ij}m_{l_j} \Big ] c_m^{(\mp)},
\nonumber
\end{eqnarray}

where
$\beta$ is $\Phi_a^{(2)}-\Phi_{15}^{(3)}$ mixing angle in MQLS model,
$tg\beta= \eta_3/\eta_2$,
$C_Q = V_{CKM}$, $C_l = U_{PMNS}$ and
$K^{L,R}_a = (A^{L,R}_{Q_a})^+ A^{L,R}_{l_a}$ are the mixing
matrices specific for the MQLS model.

\underline{Scalar gluons $F_a$ couplings to fermions:}

\begin{eqnarray}
(h^{L}_{F_1})_{ij} &=&\sqrt{ 3} \frac{1}{\eta \sin\beta}
\Big [  m_{u_i} (C_Q)_{ij} -  (K^R_1)_{ik}m_{\nu_k}(\krup{K_1^L}C_l)_{kj} \Big ],
\nonumber \\
(h^{R}_{F_1})_{ij} &=&-\sqrt{ 3} \frac{1}{\eta \sin\beta}
\Big [ (C_Q)_{ij}m_{d_i} -  (C_lK^L_2)_{ik}m_{l_k}(\krup{K_2^R})_{kj} \Big ],
\nonumber \\
(h^{L}_{1F_2})_{ij} &=&-\sqrt{ 3} \frac{1}{\eta \sin\beta}
\Big [  m_{u_i} \delta_{ij} -  (K^R_1)_{ik}m_{\nu_k}(\krup{K_1^L})_{kj} \Big ],\label{const2}
\\
(h^{R}_{2F_2})_{ij} &=&-\sqrt{ 3} \frac{1}{\eta \sin\beta}
\Big [  m_{d_i} \delta_{ij} -  (K^L_1)_{ik}m_{l_k}(\krup{K_1^R})_{kj} \Big ],
\nonumber\\
(h^{R}_{1F_2})_{ij} &=&0,
\nonumber \\
(h^{L}_{2F_2})_{ij} &=&0.\nonumber
\end{eqnarray}

\underline{The largest couplings $h\sim m_t/\eta$:}

\begin{eqnarray}
S^{(+)}_1\bar{t}\tau:\hspace{5mm}(h^{L}_+)_{33} &=&\sqrt{ 3/2} \frac{ m_{t}}{\eta \sin\beta}
(K_1^LC_l)_{33} ,
\nonumber \\
S^{(-)}_1\bar{\nu_\tau}b:\hspace{5mm}(h^{L}_-)_{33} &=&\sqrt{ 3/2} \frac{ m_{t}}{\eta \sin\beta}
 (\krup{K^R_1})_{33}  (C_Q)_{33},
 \\
S_m\bar{t}\nu_\tau:\hspace{5mm}(h^{L,R}_{1m} )_{33} &=& -\sqrt{ 3/2} \frac{ m_{t}}{\eta \sin\beta}
  (K_1^{L,R})_{33}c_m^{(\pm)},
\nonumber\\
F_1\bar{t}b:\hspace{5mm}(h^{L}_{F_1})_{33} &=&\sqrt{ 3} \frac{ m_{t}}{\eta \sin\beta}
 (C_Q)_{33},
\nonumber \\
F_2\bar{t}t:\hspace{5mm}(h^{L}_{1F_2})_{33} &=&-\sqrt{ 3} \frac{ m_{t}}{\eta \sin\beta}.
\nonumber
\label{eq:hfqq}
\end{eqnarray}

$m_t/\eta\sim 0.7!$

\section{$t\bar{t}$ Production at the Tevatron}
With account these large couplings of scalars with t-quarks,
scalar leptoquarks and scalar gluons may give significant contribution in $t\bar{t}$-quark production at Tevatron.

The latest CDF data on cross section
and forward-backward asymmetry of the $t\bar{t}$ production at
the Tevatron CDF \cite{CDF9913,CDFAFB2009}
\begin{eqnarray}
\sigma_{t\bar{t}} & = & 7.5 \pm 0.31 (stat) \pm 0.34 (syst) \pm 0.15 (lumi) \mathrm{pb}, \,
\label{expcspptt09}
\\
A_{\rm FB}^{t \bar t} & = & 0.193 \pm 0.065~(\rm{stat}) \pm 0.024~(\rm{sys}).\,
\label{AFBpptt09}
\end{eqnarray}
\underline{$\sigma_{t\bar{t}}$~SM~prediction}~{\small\cite{Cacciari:2008zb}}:
\begin{eqnarray}
\sigma_{t\bar{t}}^{SM} & = & 7.35 {~}^{+0.38}_{-0.80}~\mathrm{(scale)}
{~}^{+0.49}_{-0.34} ~\mathrm{(PDFs)} \mathrm{[CTEQ6.5]}  \, \mathrm{pb}  \div
\label{sppttSM}
\\
\notag &\phantom{\div}& 7.93 {~}^{+0.34}_{-0.56}~\mathrm{(scale)}
{~}^{+0.24}_{-0.20} ~\mathrm{(PDFs)} \mathrm{[MRST2006nnlo]} \, \mathrm{pb}.
\end{eqnarray}
\underline{$A_{\rm FB}^{t \bar t}$ SM prediction} {\small\cite{antunano-2007}}:
\begin{equation}
   A_{\rm FB}^{t \bar t} = 0.051(6),
\end{equation}
\begin{equation}
    A_{\rm FB}^{t \bar t} = \frac{N_t (\cos \theta >0)-N_t (\cos \theta <0)}
{N_t(\cos \theta >0)+N_t(\cos \theta <0)}.
\end{equation}

The measured at CDF forward-backward asymmetry has significant ($\approx 2 \sigma$) deviation from predictions \cite{antunano-2007}. This may be indication of new physics.

\begin{figure}[htb]
\begin{center}
\includegraphics[scale=0.45,keepaspectratio=true]{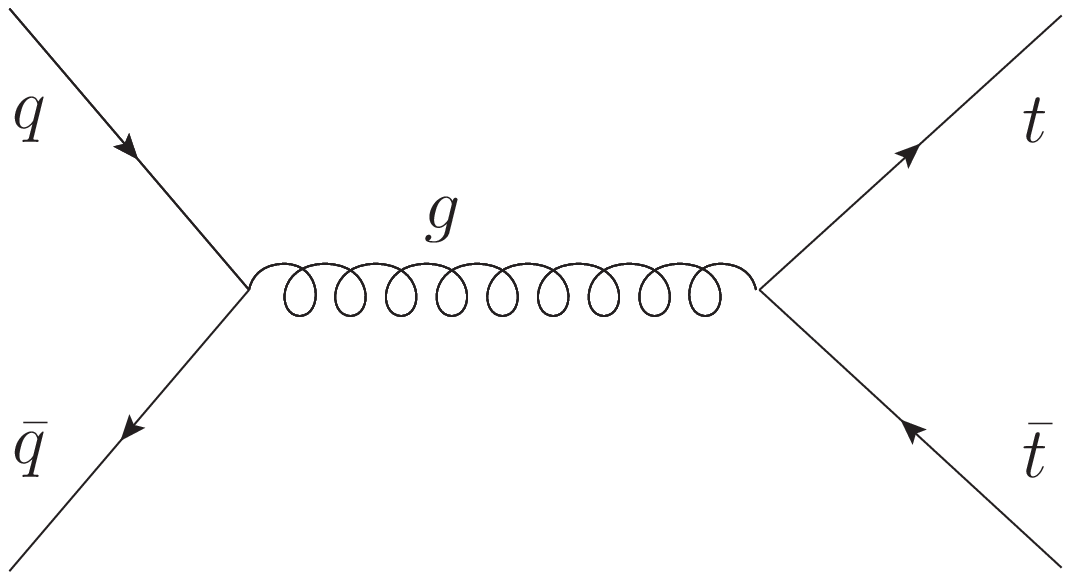}\\
\includegraphics[scale=0.55,keepaspectratio=true]{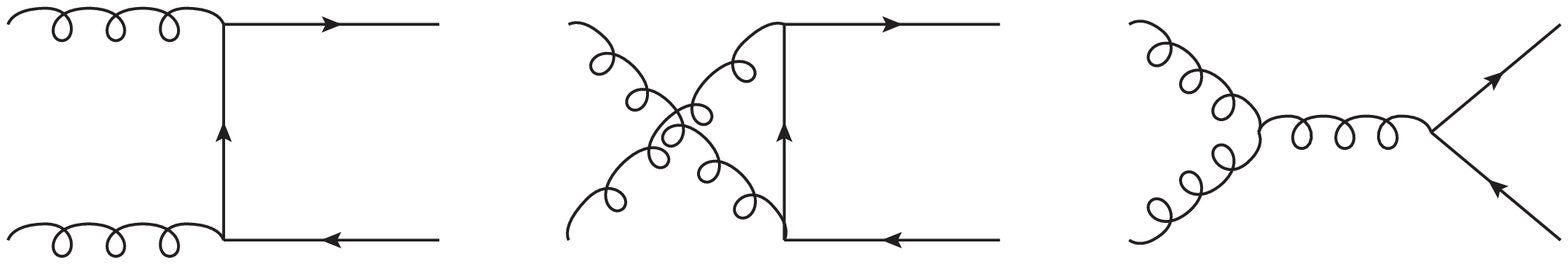}
\label{gg_tt}
\caption{Partonic subprocesses $q\bar{q}\rightarrow t\bar{t}$, $gg\rightarrow t\bar{t}$}
\end{center}
\end{figure}

The LO parton subprocesses of  $p\bar{p}\rightarrow t\bar{t}$ in SM are described
by diagrams at Fig. \ref{gg_tt} of order $\alpha_s^2$.


The well-known $p\bar{p}\rightarrow t\bar{t}$ LO cross sections have form
\begin{eqnarray}
\frac{ d\sigma(q\bar{q}\to t \bar{t}) }{d\cos \hat{\theta}} &=&
 \frac{\alpha_s^2\pi \beta}{9\hat{s}}
\left ( 1+\beta^2 c^2+ 4m_t^2/\hat{s}
\right ),\\ \nonumber
\sigma(q\bar{q}\to t \bar{t}) &=& \frac{4\pi \alpha_s^2\beta}{27\hat{s}}
\left (
3-\beta^2
\right ),
\\
\frac{d\sigma(gg\to t \bar{t})}{d\cos \hat{\theta}} &=&
\alpha_s^2 \: \frac{\pi \beta}{6 \hat{s}}
\left(\frac{1}{1-\beta^2c^2}-\frac{9}{16}\right)
\left(1 + \beta^2 c^2 +2(1-\beta^2)-\frac{2 (1-\beta^2)^2}{1-\beta^2 c^2}\right),
\\ \nonumber
\sigma(gg\to t \bar{t}) &=& \frac{\pi  \alpha_s^2 }{48 \hat{s}}
\left[
\left(\beta ^4-18 \beta ^2+33\right) \log \left(\frac{1+\beta }{1-\beta }\right)+
\beta  \left( 31 \beta ^2-59 \right)
\right],
\end{eqnarray}

where $c = \cos \hat{\theta}$,
$\hat{\theta}$ is the  scattering angle of $t$-quark in the parton center of mass frame,
$\hat{s}$ is the invariant mass of $t \bar{t}$ system, $\beta = \sqrt{1-4m_t^2/\hat{s}}$.

No sources of order $\alpha_s^2$ for the forward-backward asymmetry.

\vspace{2mm}
\underline{MQLS model contributions in $t\bar{t}$ production}
\vspace{2mm}

In MQLS there are three kind of contributions in $t\bar{t}$-production.

\begin{enumerate}
  \item Z' tree s-channel process,
  \item Scalar gluons tree processes,
  \item 1-loop $gt\bar{t}$ effective vertex.
\end{enumerate}

\underline{Z' tree s-channel process}

\begin{figure}[htb]
\begin{center}
  \includegraphics[scale=0.5,keepaspectratio=true]{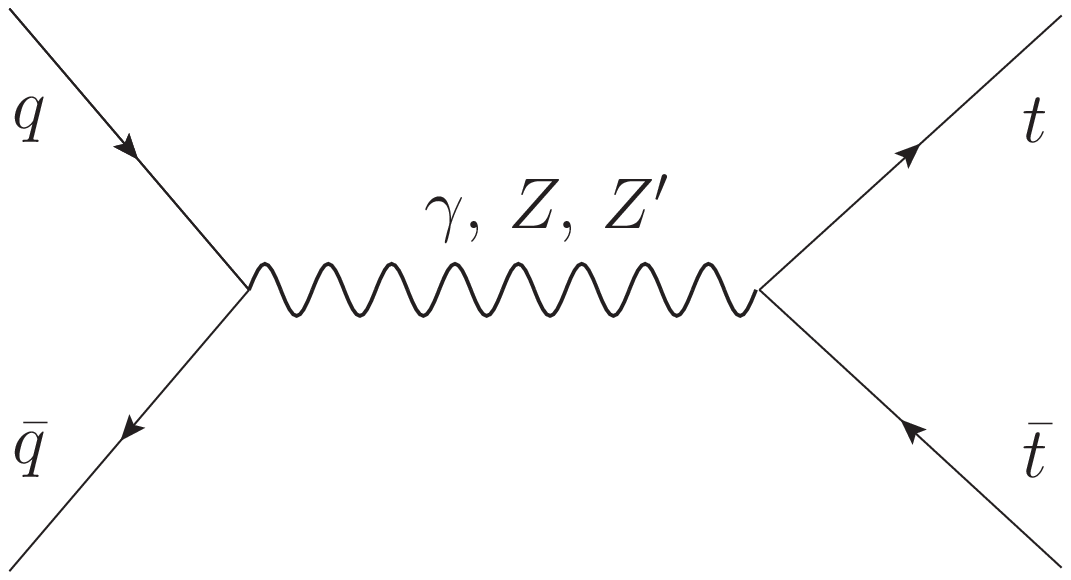}
\caption{Subprocess $q\bar{q}\stackrel{\,\gamma,\,Z,\,Z'}{\rightarrow} t \bar{t}$}
\label{qq_Z_tt}
\end{center}
\end{figure}

Partonic subprocess $q\bar{q}\stackrel{\,\gamma,\,Z,\,Z'}{\rightarrow} t \bar{t}$ is pictured at Fig. \ref{qq_Z_tt}.
Because initial quarks have singlet color state these diagrams do not interfere with octet state QCD tree processes.

We obtain differential cross section of $q\bar{q}\stackrel{\,\gamma,\,Z,\,Z'}{\rightarrow} t \bar{t}$ with account masses of final $t$-quarks in the form
\begin{eqnarray}
\frac{ d\sigma(q\bar{q}\stackrel{\,\gamma,\,Z,\,Z'}{\rightarrow} t \bar{t}) }{d\cos \hat{\theta}}=\frac{\pi\alpha_{em}^2 \hat{s}\beta}{2}\sum_{i,j=\gamma,Z,Z'} K_{ij}\mathrm{Re}(P_i(\hat{s})P_j^*(\hat{s})),
\end{eqnarray}
$\cos \hat{\theta}\equiv c$.

Here,
\begin{eqnarray}
\nonumber
K_{ij}&=&A_{ij}\left(2+\beta^2(c^2-1)\right)+B_{ij}\beta^2(c^2+1)+2C_{ij}\beta c,\\\nonumber
A_{ij}&=&(a^q_ia^q_j+v^q_iv^q_j)v^t_iv^t_j,\\
B_{ij}&=&(a^q_ia^q_j+v^q_iv^q_j)a^t_ia^t_j,\\\nonumber
C_{ij}&=&(a^q_iv^q_j+v^q_ia^q_j)(a^t_iv^t_j+v^t_ia^t_j),\\\nonumber
P_i(\hat{s})&=&\frac{1}{\hat{s}-M_i^2+iM_i\Gamma_i},
\end{eqnarray}
$v^q_i$, $a^q_i$ -- vector and axial-vector couplings of $q$-quark with $i$-th neutral boson.

For $M_{Z'}>1.4$ TeV (current experimental limit \cite{PDG2008,Smirnov:2009ed}),
contributions of $Z'$ to cross section and FB asymmetry of the $t \bar{t}$ production is small due smallness of couplings
\begin{eqnarray}
  \Delta \sigma(p\bar{p}\to t\bar{t})&\sim& +0.05\div 0.1\,\mbox{pb},\\
  \Delta A_{FB}^{t\bar{t}}&\sim& +0.003.
\end{eqnarray}

\underline{Scalar gluons tree processes}

\begin{figure}[htb]
\begin{center}
\includegraphics[scale=0.5,keepaspectratio=true]{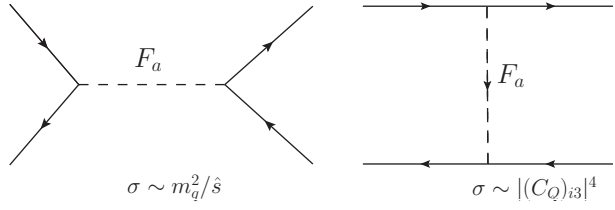}
\label{Ftt}
\caption{$s$- and $t$-channel diagrams of $q\bar{q}\to F_a\to t\bar{t}$.}
\end{center}
\end{figure}

Contributions of diagrams at Fig. 4
 are suppressed by factors $m_q^2/\hat{s}$ or $|(V_{CKM})_{i3}|^4$

\begin{eqnarray}
  \Delta \sigma(p\bar{p}\to t\bar{t})&\sim& 0.0001\,\mbox{pb},\\
  \Delta A_{FB}^{t\bar{t}}&\sim& 10^{-6}.
\end{eqnarray}

\underline{1-loop $gt\bar{t}$ effective vertex}

\begin{figure}[htb]
\begin{center}
\includegraphics[scale=0.7,keepaspectratio=true]{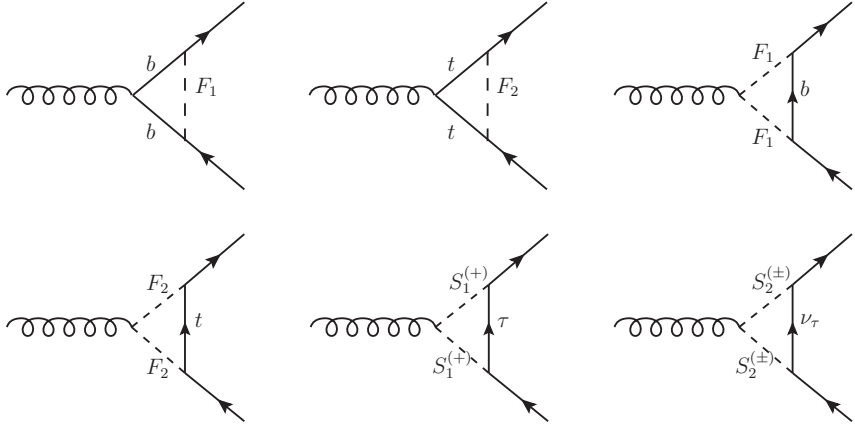}
\caption{1-loop main contributions into effective $gt\bar{t}$-vertex in MQLS-model.}
\label{gtt_effvertex}
\end{center}
\end{figure}


The significant contributions to $t\bar{t}$ production  may arise from loop corrections to the $gt\bar{t}$-vertex.

Following the parametrization in Ref.~\cite{Stange:1993td,Cao:2008qd}, the
effective matrix element of $gt\bar{t}$, including the one-loop corrections,
can be written as
\begin{equation}
-ig_{s}T^{a}\bar{u}_{t}\Gamma^{\mu}v_{\bar{t}},\label{eq:GenForm0}
\end{equation}
with
\begin{equation}
\Gamma^{\mu}=(1+\alpha)\gamma^{\mu}+i\beta\sigma^{\mu\nu}q_{\nu}+\xi\left(\gamma^{\mu}
-\frac{2m_{t}}{\hat{s}}q^{\mu}\right)\gamma_{5}.
\label{eq:GenForm}
\end{equation}
 where the loop-induced form factors $\alpha$, $\beta$ and $\xi$
are usually refereed as the chromo-charge, chromo-magnetic-dipole and chromo-anapole, respectively. Here, $g_{s}$ is the strong coupling
strength, $T^{a}$ are the color generators, $q=p_{t}+p_{\bar{t}}$,
and $\hat{s}=q^{2}$. After summing over the final state and averaging
over the initial state colors and spins, the partonic total cross
section of $q\bar{q}\to g\to t\bar{t}$ is~\cite{Stange:1993td}
\begin{equation}
\hat{\sigma}=\frac{8\pi\alpha_{s}^{2}}{27\hat{s}^{2}}\sqrt{1-\frac{4m_{t}^{2}}{\hat{s}}}\,
\biggl\{\hat{s}+2m_{t}^{2}+2\mathrm{Re}\left[(\hat{s}+2m_{t}^{2})\alpha+3m_{t}\hat{s}\,\beta\right]
\biggr\},\,\label{eq:Hardxsect}
\end{equation}
 where $\alpha_{s}\equiv g_{s}^{2}/(4\pi)$, and $\mathrm{Re}$ denotes taking
its real part.
In MQLS-model main 1-loop contributions into effective $gt\bar{t}$-vertex  are described by diagrams at Fig.\ref{gtt_effvertex}.
The parameters $\alpha$, $\beta$ can be calculated using the diagrams shown in Fig. \ref{gtt_effvertex}
and  the coupling constants (\ref{const1}-\ref{const2}).

\section{Summary}

\begin{itemize}
  \item   The contributions
to the cross section $\sigma_{t\bar{t}}$ and to the forward-backward asymmetry
$A_{\rm FB}^{t \bar t}$ of $t\bar{t}$ production at the Tevatron from new $Z'$,$S^{(\pm)}_a$, $F_a$ particles predicted by the MQLS-model
are calculated.
\item These contributions in tree approximation  are shown to be small ($\Delta\sigma~\sim~ 0.1$~pb, $\Delta A_{\rm FB}^{t \bar t}\sim 0.003$).
\item The scalar doublets $S^{(\pm)}_a$, $F_a$ may give the significant contributions to the 1-loop $gt\bar{t}$ effective vertex.
\item The lower mass limits for scalar gluons  $$m_F \gtrsim 320\, GeV$$  are obtained from the data on direct searches at Tevatron.
\item At $m_{F_1}\lesssim 990\,GeV$ the scalar gluon $F_1$ can be evident at LHC at the significance not less that $3\sigma$ (for $L=10\,fb^{-1}$).
\end{itemize}

\underline{Acknowledgement}

The work is supported by the Ministry of Education and Science of Russia
under state contract No.P2496
of the Federal Programme "Scientific and Pedagogical Personnel of Innovation Russia"
for 2009-2013 years.


\end{document}